\documentclass[showpacs,twocolumn]{revtex4}
\usepackage{amsfonts,amssymb,amsbsy,bm,graphicx}

\def\h{\hskip 7mm}
\def\que#1#2{\displaystyle\frac{#1}{#2}}

\let\De=\Delta
\let\al=\alpha

\def\sen{\mathop{\rm sen}\nolimits}
\let\ra=\rightarrow
\let\va=\varphi

\let\la=\lambda

\begin{document}


\title{On the magnitude of the energy flow inherent in
zero-point radiation}

\author{Rafael Alvargonz\'alez}
\affiliation{Universidad Complutense, 28040 Madrid, Spain}

\date{\today}

\begin{abstract}
The spectrum of zero-point radiation is relativistically invariant
and its spectral density function is therefore inversely
proportional to the cubes of its wavelengths. For its energy to be
finite, there must exist a minimum wavelength, $q_\lambda$. The
measurements of the apparent attraction between two uncharged
conductor plates, placed in a vacuum at a temperature close to
absolute zero, made by Sparnaay in 1958 allow us to deduce that
the energy flow of the zero-point radiation which comes of or into
an area $(q_\lambda)^2$, corresponds with the emission of one
photon of wavelength $q_\lambda$ per $q_\tau$\
$(q_\tau=q_\lambda/c)$, plus one photon of wavelength $2q_\lambda$
per $2^3q_\tau$, etc., up to one photon of wavelength $nq_\lambda$
per $n^3q_\tau$. This energy flow is enormous, but Sparnaay's
experiments implied only photons whose wavelengths were greater
than $5\times10^{-5}$ cm, and zero-point radiation may include
only photons with wavelengths greater than $xq_\lambda$, being $x$
an integer, perhaps very great.
\end{abstract}

\maketitle

\section{Introduction}

Sparnaay's 1958~\cite{Spa58} experiments exposed the existence of
zero-point ra\-dia\-tion, which Nernst had considered as a
possibility in 1916. However, this was not Sparnaay's intention,
as he was only trying to check Casimir's hy\-po\-the\-sis about
the mutual attraction of two uncharged conductor plates placed
very close together in a vacuum~\cite{Cas49}. This attraction must
disappear when the temperature approaches absolute zero, but
Sparnaay found that at that temperature there was still some
attraction not accounted for by Casimir's hypothesis, being
independent of temperature and obeying a very simple law, it is
directly proportional to the surface area of the plates, and
inversely proportional to the 4$^{\rm th}$ power of the distance
``$d$"\ between them. Sparnaay observed this force when the plates
were located in a near-perfect vacuum at near-zero absolute
temperature. For a distance of $5\times10^{-5}$~cm between the
plates, he was able to measure a force of 0.196 g \ cm \ s$^{-2}$,
and deduced the formula
$$f=\que{k_s}{d^4};\qquad\ \hbox{where}\ \ k_s=1.3\times10^{-18}\ {\rm
erg} \ {\rm cm}$$

In a near-perfect vacuum and at near absolute zero temperature, which means
the absence of any ``photon gas", the phenomenon observed by Sparnaay could
only be produced by a radiation inherent in space. This could only be the case
if its spectrum is relativistically invariant, which could only happen if its
spectral density function is inversely proportional to the cubes of the
wavelengths. In other words, if the number of photons of wavelength $\la$,
which strike a given area within a given time is inversely proportional to
$\la^3$.

A function of spectral density inversely proportional to the cubes
of the wavelengths, implies a distribution of energies which is
inversely proportional to the 4$^{\rm th}$ power of the
wavelengths, because the photons energies are inversely
proportional to their wavelengths. In 1969, Timothy H.
Boyer~\cite{Boy69} showed that the spectral density function of
zero-point radiation is
\begin{equation}
f_\va(\la)=\que1{2\pi^2} \que1{\la^3_*};
\end{equation}
where $\la_*$ is the number giving the measurement of wavelength
$\la$. This function produces the next for the corresponding
energies
\begin{equation}
E_\va(\la)=\que1{2\pi^2} \que{hc}\la \que1{(\la_*)^3} .
\end{equation}

For $\la\ra0$, $E_\va(\la)\ra\infty$. There must therefore be a
threshold for $\la$, which will hereafter be designated by the
symbol $q_\lambda$.

In the Sparnaay effect, the presence of a force which is inversely
proportional to the 4$^{\rm th}$ power of the distance ``$d$"\ between the
plates, while the photons distribution of energies is inversely proportional
to the 4$^{\rm th}$ power of the wavelengths, leads us to infer that the cause
of the apparent attraction must lie in some factor which varies inversely with
wavelength and which produces a different behaviour in photons of wavelength
equal to or greater than ``$d$", since:
$$\sum^\infty_{n=1}kn^{-5}\ \ra\ \que k{4d^4}$$

The factor in question turns out to be the proportion of photons which are
reflected, not absorbed; that is the coefficient of reflection $\rho$, while
the different behaviour is caused by the obstacle which each plate offers to
the reflection by the inner plate of the other one, of photons of wavelengths
equal to or greater than ``$d$". As we know, the energy which in transferred
by a reflected photon is double that transferred by one which is absorbed.

\section{Interpretation of the experimental results}

We now come to the title of this paper, an attempt to find the magnitude of
the energy flow which is inherent in zero-point radiation; two radiations with
the same spectrum, i.e. with the same energy distribution by wavelength, can
have very different energy flows.

The results of Sparnaay's experiments provide a link with reality
which is enough to show the intensity of the energy flow belonging
to zero-point radiation. So, remembering the requirement that in
zero-point radiation there must exist a threshold, $q_\lambda$, for
the wavelength of the photons, we will now test the hypothesis
that the results agree with an energy flow which would produce the
incidence in a hypothetical area $(q_\lambda)^2$ of:
$$\begin{array}{lrll}
\hbox{One photon of wavelength} & q_\lambda, & \hbox{per} & x
q_\tau\
(q_\tau=q_\lambda/c)\\[+7pt]
\hbox{One photon of wavelength} & 2q_\lambda, & \hbox{per} & x 2^3q_\tau
\ldots \\[+7pt]
\hbox{One photon of wavelength} & nq_\lambda, & \hbox{per} & x
n^3q_\lambda
\end{array}$$

This radiation implies an energy flow of:
\begin{eqnarray}
W_\va & = & \que1x \que{hc}{q_\lambda q_\tau}\
\left(1+\que1{2^4}+\cdots+\que1{n^4}\right) \nonumber \\
& = & \que1x \que{\pi^4}{90}
\que{hc}{q_\lambda q_\tau} \quad
\hbox{per} \ (q_\lambda)^2 .
\end{eqnarray}

From Sparnaay's experiments and our earlier argument, this energy flow should
cause the difference between the flows transmitted to the outer faces, and
those transmitted to their inner faces to be:
\begin{eqnarray}
\Delta W_\lambda & =  & \rho \que{hc}{q_\lambda q_\tau} \que1x
\left [ \que1{n^5}+ \que1{(n+1)^5}+ \cdots + \que1{n^5} \right ]_{m \ra\infty}
\nonumber \\
&  & \nonumber \\
& & \hbox{per} \ q_\lambda\ \hbox{of surface}
\end{eqnarray}
where $n$ is the distance between the plates expressed in $q_\lambda$.
Hence
\begin{eqnarray}
\Delta W_\lambda &  = & \rho \que{hc}{q_\lambda q_\tau} \que1x \que1{4n^4}
\quad
\hbox{per}\ (q_\lambda)^2 .
\end{eqnarray}

To simplify the next stages of reasoning we will use the system of measurement
($e$, $m_c$, $c$), in which the basic magnitudes are the quantum of electric
charge $e$, the electron mass $m_e$, and the speed of light $c$. In this
system the units of length and time are respectively: $l_e=e^2/m_ec^2$,
$t_e=e^2/m_ec^3$; producing $h=\que{2\pi}\al \que{m_el_e^2}{t_e}$;
$c=\que{l_e}{t_e}$.

Clearly, if $l_e=k_\lambda q_\lambda$, we have
$t_e=l_e/c=k_\lambda q_\lambda/c=k_\lambda q_\tau$. On the other
hand if $n$ is the number representing the measurement of the
distance between the plates in $q_\lambda$, the number which
represents it in $l_e$ is $D_{l_e}$, $D_{l_e}k$ being equal to
$n$. It must be remembered that $k_\lambda$ is certainly a very
large number; there have been detected cosmic rays photons with
wavelengths up to $10^{20}$ eV, which would imply $k_\lambda$
values of the order of $2.273\times10^{11}$. The true value of
$k_\lambda$ could be much greater than this.

If in (5) we substitute $q_\lambda$ by $l_e/k_\lambda$; $q_\tau$
by $t_e/k_\lambda$; $n$ by $D_{l_e}k_\lambda$, and $h$ by
$\que{2\pi}\al \que{m_el_e^2}{t_e}$, and remember that the flow
per $(l_e)^2$ is $(k_\lambda)^2$ times greater than the flow per
$(q_\lambda)^2$, we can write
$$\De W_{(e,m_e,c)} = \que\rho x \que{2\pi}\al \que{m_el_e^2}{t_e}
\que1{(l_e/k_\lambda)(t_e/k_\lambda)}$$
$$ \times \que1{4D_{l_e}^4(k_\lambda)^4}(k_\lambda)^2\
\ \hbox{per}\ (l_e)^2$$
That is
\begin{equation}
\De W_{(e,m_e,c)}=\que{\rho\pi}{2\al x} \que1{D^4_{l_e}} \que{m_ec^2}{t_e}\
\quad
\hbox{per}\ (l_e)^2 .
\end{equation}

In (6) we find that the factor $k_\lambda$ has disappeared. If
this were not so, we would arrive at absurd results, since the
value of $\rho$ would be provided by:
$$\rho=\que{\De W_{(e,m_e,c)}}{m_e c^2/t_e}\que{2\al x}\pi
D_{l_e}^4\,(k_\lambda)^a$$

For $a>0$ $\rho$ would be huge and for $a<0$ $\rho$ would be tiny;
it must be remembered that $k_\lambda$ must be greater than
$2.273\times10^{11}$. The values of $\rho$ must always be less
that 1, and for a radiation where photons predominate with
wavelengths slightly greater than $5\times10^{-5}$ cm, these
values must be greater than 0.4.

Since we must consider the energy flows for both the plates, we find that they
produce an apparent force of attraction which can be expressed by:
\begin{equation}
F=\que{2 \Delta W_{(e,m_e,c)}}c=\que{\rho\pi}{\al x} \que1{D_{l_e}^4}
\que{m_el_e}{t_e^2}\ \ \hbox{per}\ (l_e)^2 .
\end{equation}

In order to express (7) in terms of g \ s per cm$^2$, we have to include the
transformation coefficients 1 cm$=k_{l_e} l_e$; 1 s$=k_{t_e} t_e$;
1 g$=k_{m_e} m_e$, remembering to multiply by $(k_{l_e})^2$ in order to move
to forces per cm$^2$, and that $D_{l_e}=k_{l_e}d$ cm, from which we obtain the
following force per cm$^2$:
\begin{equation}
F_{\rm (c,g,s)}=\que{\rho\pi}{\al x} \que{(k_{t_e})^2}{(k_{l_e})^3k_{m_e}}
\que1{d^4} \que{{\rm g} \ {\rm cm}}{{\rm s}^2} .
\end{equation}

Since $k_{l_e}=3.548692\times10^{12}$;
$k_{t_e}=1.063871\times10^{23}$; $k_{m_e}=1.097668\times10^{27}$;
$d=5\times10^{-5}$; by inserting these values in (8), we reach
\begin{equation}
F_{\rm (c,g,s)}=\que\rho x 15.893023\ \que{{\rm g} \ {\rm
cm}}{{\rm s}^2}
\end{equation}

Zero-point radiation comes from all directions within the half of space which
is faced by the outer side of each plate, and its normal component is given,
for each direction, by $F\sen\al$, where $\al$ is the angle between the plane
of the plate and the direction in question. If we imagine a sphere of radius
$F$, it is not hard to see that the number with measures the sum of the normal
components is the number which measures the volume of the hemisphere of radius
$F_*$, where $F_*$ is the number measuring $F$ in the system of units employed.

However $\De W_\lambda$, as shown in (5), is not the flow which
arrives from each and every direction in the said half-space, but
the sum of all the flows which arrive from all those directions.
In order to make the necessary correction we have only to divide
the number $\que{2\pi}3(F_*)^3$ by $2\pi(F_*)^2$, which represents
the area of the hemisphere of radius $F_*$, this amounts to
dividing the total sum between the total of the directions to be
considered.  The resulting value is
\begin{equation}
F_0=\que13F_{\rm (c,g,s)}=\que\rho x 5.297674\ \que{{\rm g} \ {\rm
cm}}{{\rm s}^2}
\end{equation}

However, this relation, which makes it easier to visualize the problem, is not
the correct one, because the radiation which strikes at an angle $\va$ from
the vertical has a trajectory between the plates which is given by
$d/\cos\va$ (Fig.~1).

\begin{figure}
\centering
\resizebox{0.75\columnwidth}{!}{\includegraphics{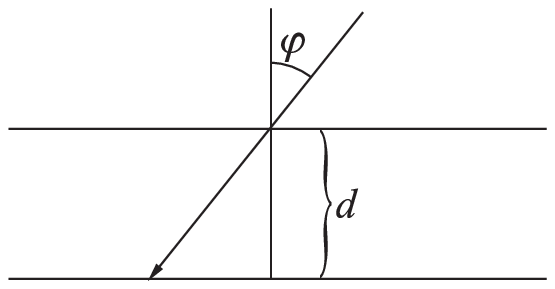}}
\caption{}
\end{figure}

Therefore the value of $\De W_\lambda$ to be considered is:
\begin{eqnarray}
\De W_{\la\va} & = & \que{hc}{q_\lambda q_\tau} \que\rho x \left [
\que1{n^5/\cos^5\va}+\que1{(n/\cos\va+1)^5}+ \cdots \right . \nonumber \\
& + & \left . \que1{n/\cos\va+m)^5} \right ] ,
\end{eqnarray}
replacing (4), so that we obtain:
$$\De W_{\la\va}=\que\rho x \que{hc}{q_\lambda q_\tau} \que14
\que1{n^4/\cos^4\va},$$
from which we reach:
\begin{equation}
F_{\va(e,m_e,c)}=\que{\rho\pi}{\al x} \que{\cos^4\va}{D^4}
\que{m_el_e}{t^2_e}\ \ \hbox{per}\ (l_e)^2,
\end{equation}
replacing (7).

This force is equal to $F_{(e,m_e,c)}\cos^4\va$ and its component normal to the
plane of the plates is $F_{(e,m_e,c)}\cos^5\va$. On fig. 2 it can be seen that
to each point $P$ on the circumference of radius $F$ there corresponds a point
$P_\va$ whose coordinates are those of $P$ multiplied by $\cos^4\va$. The
points $P_\va$ form a line within the circle, and the sum of the component
normal to the plates is now given by the volume produced by the area enclosed
within that line and by the axis of coordinates when it turns around the axis
$OY$, instead of the volume produced by the quadrant of circle turning around
that axis. The volume in question is given by the product of the said area and
the circle described by its center of gravity~\cite{Rou49} (Fig. 2).

\begin{figure}
\centering
\resizebox{0.75\columnwidth}{!}{\includegraphics{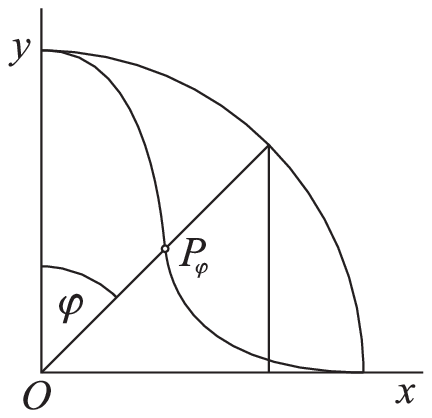}}
\caption{}
\end{figure}

The parametric coordinates of the curve thus described are
$$x=F_*\cos^4\va\,\sen\va\qquad\ y=F_*\cos^5\va$$

To the ordinate $F_*\cos^5\va$ there corresponds the differential element of
abscissa $F_* (\cos^5\va-4\cos^3\va\,\sen^2\va )$. Therefore
$$S=(F_*)^2\int^{\pi/2}_0 ( \cos^{10}\va-4\cos^8\va\,\sen^2 \va ) d\va;$$
whence:
$$S=0.214757(F_*)^2$$

The circle described by the center of gravity as it turns round axis $OY$ is
one whose radius is the abscissa of that center of gravity, which is given by:
$$x_G  S=\int^{\pi/2}_0\que x2  x\,dy=\que52(F_*)^3\int^{\pi/2}_0
(\cos^{12}\va\,\sen^3\va ) \,d\va;$$
whence:
$$x_G  S=(F_*)^3 \que1{39}\ \ \ \hbox{and}\ \ \ x_G=0.119395F_*$$

The volume to be considered is therefore
$$V=(F_*)^3\times0.214757\times2\pi\times0.119395=0.1611(F_*)^3$$

While the force to be considered is
$$F=\que{0.1611(F_*)^3}{2\pi(F_*)^2}=0.0256402F_*$$

With the value of $F_*$ given by (9)
$$F=\que\rho x 0.407513\ \que{{\rm g} \ {\rm cm}} {{\rm s}^2}.$$
Using this with the results of Sparnaay's measurements, we obtain
$$\que\rho x 0.407513=0.196 ;\ \ \ \hbox{whence}\ \rho=0.481 x$$

Since $x$ must be a natural number, and $\rho$ must fall between 0
and 1 the only possible results are
$$\begin{array}{ll}
x=1 & \rho=0.481 \\[+4pt]
x=2 & \rho=0.962 \end{array}$$

The second of these pairs is unlikely for wavelengths close to
$5\times10^{-5}$ cm. We can therefore deduce that the energy flow
of zero-point radiation is the results of the incidence, in an
area $(q_\lambda)^2$, of:
$$\begin{array}{lrlr}
\hbox{One photon of wavelength} & q_\lambda & \hbox{per} & q_\tau\\[+7pt]
\hbox{One photon of wavelength} & 2q_\lambda, & \hbox{per} & 2^3q_\tau \ldots
\\[+7pt]
\hbox{One photon of wavelength} & nq_\lambda, & \hbox{per} &
n^3q_\tau
\end{array}$$

\section{Conclusions}

\h Returning to (3);
$W_\lambda=\que{\pi^4}{90x} \que{hc}{q_\lambda q_\tau}$ per
$(q_\lambda)^2$, and expressing it through the $(e,m_e,c)$ system
and as an energy flow by $(l_e)^2$, we obtain
$$W_{(e,m_e,c)}=\que{\pi^4}{90x} \que{2\pi}\al \que{m_el_e^2}{t_e}
c \que{(k_\lambda)^2}{(l_e/k_\lambda)(t_e/k_\lambda)}$$

Whence, where $x=1$,
$$W_{(e,m_e,c)}=\que{\pi^4}{45\al}\,(k_\lambda)^4\,\que{m_ec^2}{t_e};$$
and a force
$$F_{(e,m_e,c)}=\que{\pi^5}{45\al}\,(k_\lambda)^4\,\que{m_ec}{t_e^2},$$
over an area $(l_e)^2$.

This force, expressed in the (c,g,s) system and over 1 cm$^2$ is produced by
\begin{eqnarray}
F_{\rm (c,g,s)} & = & \que{\pi^5}{45\al}\,(k_\lambda)^4\
\que{({\rm g}/k_{m_e})({\rm cm}/k_{l_e})}{({\rm
s}/k_{t_e})^2}\times
(k_{l_e})^2 \nonumber \\
& = & 3.409994\times10^{34}(k_\lambda)^4\ \que{{\rm g} \ {\rm
cm}}{{\rm s}^2}
\end{eqnarray}

As before, the normal force per cm$^2$ of surface is given by the third part of
this value (see (10))
$$\que{F_{\rm (c,g,s)}}3=1.136647\times10^{34}(k_\lambda)^4\
\que{{\rm g} \ {\rm cm}} {\rm s}$$
per cm$^2$ of surface.

Given the importance of $k_\lambda$, this force is huge and, at
first sight, impossible to imagine. However, the rigorous equality
prevailing between the energy flows arriving from all directions
of space implies that the results are systematically nil. The
presence of this force can only be detected when its equality of
arrival is disturbed, as a result of phenomena such as the
Sparnaay effect, or through the interaction of the photons with
free elemental particles. Ob\-vious\-ly Sparnaay's experiments
implied only photons whose wavelengths were greater than
$5\times10^{-5}$ cm, and zero-point radiation may include only
photons with wavelengths greater than $xq_\lambda$, being $x$ an
integer perhaps very great. Nonetheless the fact that this force
may be immensely great is a surprising datum in the reality of the
Universe. A surprising datum from which surprising consequences
may derive.


\begin{thebibliography}{99}

\bibitem{Spa58}
M. J. Sparnaay, Physica {\bf 24}, 51 (1958).

\bibitem{Cas49}
H. B. J. Casimir, J. Chim. Phys. \textbf{46}, 407 (1949).

\bibitem{Boy69}
T. H. Boyer, Phys. Rev. \textbf{182}, 1374 (1969);
Sci. Am. N. 10, 42 (1985).

\bibitem{Rou49}
E. Rouch\'e and Ch. Comberousse,{\em Trait\'e de Geometrie}
Vol. 2 (Gauthier-Villars, Paris, 1949)  p. 228.

\bibitem{Mil94}
P. W. Milonni, {\em The Quantum Vacuum}  (Acad. Press, New York, 1994) pp. 275-280.

\bibitem{Alv00}
R. Alvargonz\'alez, Rev. Esp. Fis. \textbf{14} (4), 32 (2000).
\end{thebibliography}
\end{document}